\documentclass[journal]{IEEEtran}

\usepackage{cite}
\usepackage{comment}
\ifCLASSINFOpdf
  \usepackage[pdftex]{graphicx}
\else
    \usepackage[dvips]{graphicx}
\fi
\usepackage{multicol}
\usepackage{multirow}

\usepackage{hyperref}

\usepackage{amsmath}
\usepackage{algorithm}
\usepackage{algpseudocode}
\usepackage{listings}
\usepackage{amssymb}

\usepackage{pifont}
\newcommand{\xmark}{\ding{55}}%
\usepackage{xcolor}

\usepackage{tcolorbox}
\usepackage{rotating}
\usepackage{textpos}

\definecolor{codegreen}{rgb}{0,0.6,0}
\definecolor{codegray}{rgb}{0.5,0.5,0.5}
\definecolor{codepurple}{rgb}{0.58,0,0.82}
\definecolor{backcolour}{rgb}{0.95,0.95,0.92}
\lstdefinestyle{cppstyle}{
    language=C++,
    backgroundcolor=\color{backcolour},   
    commentstyle=\color{codegreen},
    keywordstyle=\color{magenta},
    numberstyle=\tiny\color{codegray},
    stringstyle=\color{codepurple},
    basicstyle=\ttfamily\footnotesize, 
    numbers=left,
    breakatwhitespace=false,         
    breaklines=true,                 
    captionpos=b,                    
    keepspaces=true,                 
    numbersep=5pt,                  
    showspaces=false,                
    showstringspaces=false,
    showtabs=false,                  
    tabsize=2
}

\definecolor{highlight1}{RGB}{255, 204, 204}  
\definecolor{highlight2}{RGB}{255, 229, 204}  
\definecolor{highlight3}{RGB}{255, 255, 204}  
\definecolor{highlight4}{RGB}{229, 255, 204}  
\definecolor{highlight5}{RGB}{204, 255, 229}  
\definecolor{highlight6}{RGB}{204, 255, 255}  
\definecolor{highlight7}{RGB}{204, 229, 255}  
\definecolor{highlight8}{RGB}{229, 204, 255}  
\definecolor{highlight9}{RGB}{255, 204, 255}  
\definecolor{highlight10}{RGB}{192, 192, 192} 
\definecolor{red}{RGB}{200,0,0}   
\definecolor{Green}{RGB}{0,150,0} 

\usepackage{array}

\ifCLASSOPTIONcompsoc
  \usepackage[caption=false,font=normalsize,labelfont=sf,textfont=sf]{subfig}
\else
  \usepackage[caption=false,font=footnotesize]{subfig}
\fi

\usepackage{url}

\usepackage{textpos}

\begin{document}
\newcounter{obsno}
\definecolor{textHighlight}{RGB}{227,242,253}
\newcommand{\observation}[1]{
    \begin{tcolorbox}[width=\linewidth, colback=textHighlight,left=2pt,right=2pt,top=2pt,bottom=2pt]
        \textbf{{\small \MakeUppercase{Observation} }\refstepcounter{obsno}\Roman{obsno}:} {\small #1}
    \end{tcolorbox}
}

\title{A Parallel and Highly-Portable HPC Poisson Solver: Preconditioned Bi-CGSTAB with \texttt{alpaka}}

\author{\IEEEauthorblockN{Luca~Pennati\textsuperscript{*}, Måns~I.~Andersson\textsuperscript{*}, Klaus~Steiniger\textsuperscript{\textdagger}, Rene~Widera\textsuperscript{\textdagger}, Tapish~Narwal\textsuperscript{\textdagger}, Michael~Bussmann\textsuperscript{\textdagger}, Stefano~Markidis\textsuperscript{*}}
\IEEEauthorblockA{\\
\textsuperscript{*}KTH Royal Institute of Technology, Stockholm, Sweden \\
\textsuperscript{\textdagger} Helmholtz-Zentrum Dresden-Rossendorf (HZDR), Dresden, Germany}
}

\maketitle

\begin{abstract}
This paper presents the design, implementation, and performance analysis of a parallel and GPU-accelerated Poisson solver based on the Preconditioned Bi-Conjugate Gradient Stabilized (Bi-CGSTAB) method. The implementation utilizes the MPI standard for distributed-memory parallelism, while on-node computation is handled using the \texttt{alpaka} framework: this ensures both shared-memory parallelism and inherent performance portability across different hardware architectures. We evaluate the solver's performances on CPUs and GPUs (NVIDIA Hopper H100 and AMD MI250X), comparing different preconditioning strategies, including Block Jacobi and Chebyshev iteration, and analyzing the performances both at single and multi-node level. The execution efficiency is characterized with a strong scaling test and using the AMD \texttt{Omnitrace} profiling tool. Our results indicate that a communication-free preconditioner based on the Chebyshev iteration can speed up the solver by more than six times. The solver shows comparable performances across different GPU architectures, achieving a speed-up in computation up to 50 times compared to the CPU implementation. In addition, it shows a strong scaling efficiency greater than $\mathbf{90\%}$ up to 64 devices.
\end{abstract}

\begin{IEEEkeywords}
Poisson Solver; Preconditioned Bi-CGSTAB; \texttt{alpaka}; MPI Parallelization; Code Portability
\end{IEEEkeywords}

\IEEEpeerreviewmaketitle

\section{Introduction}
The solution of large linear systems is a fundamental problem in high performance computing (HPC), with applications ranging from computational fluid dynamics~\cite{gong2016nekbone}, to electromagnetics~\cite{kumar2013high}, and other scientific and engineering domains. Among these, Poisson solvers are particularly important as they arise in a wide range of problems, including electrostatics for Molecular Dynamics simulations~\cite{andersson2022breaking} and plasma simulations~\cite{kumar2013high}, and incompressible fluid flow problems~\cite{karp2024experience,burau2010picongpu}. Therefore, the development of efficient and scalable parallel linear solvers and preconditioners for the Poisson equation is a crucial task for many computational science applications.

The design of effective preconditioners is essential for improving the convergence rate of iterative solvers. Preconditioners, such as Block Jacobi and Chebyshev iteration, enhance the spectral properties of the system matrix and decrease the number of iterations needed for convergence, thereby improving the performance of the solvers. It is important that preconditioners are not only mathematically effective but also computationally efficient, being compatible with modern hardware architectures to fully utilize parallel computing resources and minimize data communication.

Together with the algorithmic challenge, we also have the complex task of supporting Poisson solvers on different architectures: achieving high performance while maintaining code portability is a significant undertaking. The rapid evolution of hardware architectures~\cite{hennessy2019new}, including multicore CPUs, manycore accelerators, and diverse GPU architectures from multiple vendors, such as NVIDIA, AMD and Intel, necessitates software frameworks that can efficiently use these resources without requiring extensive code rewrites. While traditional approaches often involve rewriting code to optimize for specific hardware, this is becoming increasingly impractical as new CPU and GPU architectures continuously arise.

To address these challenges, portability and performance-portable programming frameworks have been developed to abstract hardware-specific optimizations while enabling efficient execution across different platforms. An important example of such frameworks is \texttt{alpaka}~\footnote{alpaka GitHub: \url{https://github.com/alpaka-group/alpaka}}, which provides a single-source abstraction for parallel programming, allowing the same codebase to run efficiently on various architectures, including CPUs and GPUs from different vendors. By using \texttt{alpaka}, scientific applications can achieve high performance without giving up with maintainability and portability~\cite{alpaka2015,alpaka2016}.

In this work, we present a parallel Poisson solver based on the Preconditioned Bi-Conjugate Gradient Stabilized (Bi-CGSTAB) method~\cite{vanDerVorst1992_BiCGstab}, implemented with MPI and \texttt{alpaka} to ensure portability and high-performance. We investigate different preconditioning strategies, including Block Jacobi and Chebyshev iteration, to improve convergence. We evaluate the solver performances on both CPUs and NVIDIA and AMD GPUs, analyzing the efficiency with profiling tools such as AMD rocProf and Omnitrace. The results provide insights into the effectiveness of various preconditioning techniques and demonstrate the feasibility of writing portable and high-performance solvers for heterogeneous computing environments.

The contributions of this paper are the following:
\begin{itemize}
\item Development of a parallel Poisson solver using the Preconditioned Bi-CGSTAB method with support for multiple preconditioners, including Block Jacobi and Chebyshev iteration.
\item Implementation of a performance-portable solver using the \texttt{alpaka} framework, enabling execution across diverse hardware architectures without code modifications.
\item Usage of MPI for distributed-memory parallelism to scale across multiple compute nodes.
\item A comprehensive performance evaluation on CPU,  NVIDIA and AMD GPUs, utilizing profiling tools such as rocProf and Omnitrace to analyze efficiency and bottlenecks.
\end{itemize}

\section{Background}
\subsection{Solving the Poisson Equation}
The Poisson equation is an important partial differential equation (PDE) that arises in various scientific and engineering applications, modeling phenomena such as electrostatic potential, and gravitational fields:
\begin{equation}
    -\Delta \phi(x,y,z) = f(x,y,z) .
    \label{eq_poisson}
\end{equation}
Discretizing this equation on a grid leads to a structured linear system in the form $\mathbf{A}\boldsymbol{\phi}=\mathbf{f}$. The discrete representation of the function values is given by:
\begin{equation}
    \left.\phi(x,y,z) \right|_{x_i,y_j,z_k} = \phi_{i,j,k}, 
\end{equation}
and, employing a second order centered finite difference scheme, the second order derivative is approximated by:
\begin{equation}
    \left.- \dfrac{\partial^2 \phi}{\partial^2 x}\right|_{x_i} \approx \dfrac{-\phi_{i-1,j,k} + 2\phi_{i,j,k} - \phi_{i+1,j,k}}{\Delta x^2} .
\end{equation}
The resulting linear system can be represented by a matrix operator which depends on the boundary conditions (BCs) applied to the system. For Dirichlet boundary conditions, the discrete 1D Laplacian matrix takes the form in Eq. \ref{matrix_dirichlet}:
\begin{equation}
    \mathbf{D} = 
\begin{pmatrix}
2 & -1 & &  & \\
-1 & 2 & -1 & &  \\
 & . & . & . & \\
  &  & -1& 2 & -1 \\
 & &  &  -1 &2  
\end{pmatrix}.
\label{matrix_dirichlet}
\end{equation}
Applying second order Neumann BCs to one side of the domain yields different matrix structures, detailed in Eq. \ref{matrix_neuman}:
\begin{equation}
    \mathbf{N} = 
\begin{pmatrix}
2 & -\alpha & &  & \\
-1 & 2 & -1 & &  \\
 & . & . & . & \\
  &  & -1& 2 & -1 \\
 & &  &  -\beta &2  
\end{pmatrix},
\label{matrix_neuman}
\end{equation}
where $\alpha=2$ and $\beta=1$ or $\alpha=1$ and $\beta=2$ if Neumann BCs are applied respectively to the left or to the right of the 1D domain.

The linear system arising from the discretization of the Poisson equation over a 3D grid with dimensions $N_x$, $N_y$, $N_z$, and spatial resolution $\Delta x$,  $\Delta y$, $\Delta z$, can be expressed thanks to the Kronecker product ($\otimes$) as the matrix $\mathbf{P}$:
\begin{equation}
    \mathbf{P} = \mathbf{I}_z \otimes \mathbf{I}_y \otimes \dfrac{\mathbf{O}_x}{\Delta x^2} + \mathbf{I}_z \otimes \dfrac{\mathbf{O}_y}{\Delta y^2} \otimes \mathbf{I}_x + \dfrac{\mathbf{O}_z}{\Delta z^2} \otimes \mathbf{I}_y \otimes \mathbf{I}_x
    \label{matrix_p}
\end{equation}
where $\mathbf{I}_i$ is the identity matrix and $\mathbf{O}_i$ is the operator representing the discretized Poisson equation in 1D, namely matrix $\mathbf{D}$ or $\mathbf{N}$, defined in Eq. \ref{matrix_dirichlet} and Eq. \ref{matrix_neuman} respectively. Subscripts $x,y,z$ denote the dimension of the matrices, namely $A_i \in \Re^{N_i\times N_i}$. 

Since the linear solver implemented in this work requires the knowledge of the eigenvalues of the system, it is useful to give some details here about the eigenvalues of the matrix $\mathbf{P}$.
Thanks to the Kronecker product properties, it can be easily shown that for any square matrices $\mathbf{A}\in\Re^{r\times r}$ and $\mathbf{B}\in\Re^{s \times s}$ it holds:
\begin{equation}
\left(\mathbf{A} \otimes \mathbf{I}_s+\mathbf{I}_r \otimes \mathbf{B}\right)\left(\mathbf{u}_i \otimes \mathbf{v}_j\right) =\left(\lambda_i+\mu_j\right)\left(\mathbf{u}_i \otimes \mathbf{v}_j\right),
\label{eq_eigenKronecker}
\end{equation}
where $\lambda_i$ ($\mu_j$) are the eigenvalues and $\mathbf{u}_i$ ($\mathbf{v}_j$) the respective eigenvectors of the matrix $\mathbf{A}$ ($\mathbf{B}$).
From Eq. \ref{eq_eigenKronecker} derives that the eigenvalues and eigenvectors of the matrix $\mathbf{P}$, defined in Eq. \ref{matrix_p}, are a combination of the eigenvalues and eigenvectors of the matrices $\mathbf{O}_x$, $\mathbf{O}_y$, $\mathbf{O}_z$. More specifically, denoting with $\mu_i^x$, $\mu_j^y$ and $\mu_k^z$ the eigenvalues of matrices $\mathbf{O}_x$, $\mathbf{O}_y$, $\mathbf{O}_z$, each eigenvalue $\lambda_l$ of the matrix $\mathbf{P}$ can be written as:
\begin{equation}
    \lambda_l = \dfrac{\mu_i^x}{\Delta x^2} + \dfrac{\mu_j^y}{\Delta y^2} + \dfrac{\mu_k^z}{\Delta z^2}.
\end{equation}
An analytical expression for matrix $\mathbf{D}\in \Re^{n\times n}$ eigenvalues is available:
\begin{equation}
    \lambda_i = 4\sin^2{\left( \frac{i\pi}{2(n+1)}\right)} \quad i = 1,...,n,
    \label{eq_eigenvaluesPoissonDirichletMatrix}
\end{equation}
where $n$ is the matrix dimension. On the other hand, in the case of Neumann boundary conditions (matrix $\mathbf{N}$) there is no analytical expression for the eigenvalues. However, employing the Gerschgorin theorem~\cite{gerschgorin1931_eigenvalues}, we can estimate that the eigenvalues are between 0 and 4.

In the solver, the minimum and maximum eigenvalues of the matrix $\mathbf{P}$, $\lambda_{\min}$ and $\lambda_{\max}$, are needed. Since all eigenvalues of matrices $\mathbf{O}_i$ are positive, $\lambda_{\min}$ and $\lambda_{\max}$ can be easily computed as:
\begin{equation}
    \lambda_{\min} = \dfrac{\min(\mu_i^x)}{\Delta x^2} + \dfrac{\min(\mu_j^y)}{\Delta y^2} + \dfrac{\min(\mu_k^z)}{\Delta z^2},
    \label{eq_lambdaMin}
\end{equation}
\begin{equation}
    \lambda_{\max} = \dfrac{\max(\mu_i^x)}{\Delta x^2} + \dfrac{\max(\mu_j^y)}{\Delta y^2} + \dfrac{\max(\mu_k^z)}{\Delta z^2} .
    \label{eq_lambdaMax}
\end{equation}

\subsection{\texttt{alpaka} Library}

\texttt{alpaka}~\cite{alpaka2015,alpaka2016,alpaka2017} is a cross-platform, open-source, header-only C++20 abstraction library designed to provide portability across various computing architectures, including all the main CPUs, and GPUs on the market. \texttt{alpaka} defines and implements an abstract interface for shared-memory parallel computing and provides back-ends for CUDA, HIP, SYCL, OpenMP and other technologies. Using \texttt{alpaka} allows us to write code that runs efficiently on different hardware platforms -- what is called performance portability -- without needing to modify the underlying codebase for each specific architecture. This makes \texttt{alpaka} particularly valuable for applications, such as parallel solvers, that require significant HPC infrastructure with different kinds of accelerators and thus must be optimized across a wide range of hardware configurations.

From the programming model point of view, \texttt{alpaka} provides high-level constructs, similar to CUDA, such as threads, blocks, and grids. \texttt{alpaka} includes abstractions for managing memory on different architectures, including global and shared memories on GPUs and shared-memory parallelism on CPUs. 

In this work, leveraging \texttt{alpaka}, we can target CPUs and both NVIDIA and AMD GPUs with minimal code changes, maintaining a unified codebase. This allows us to focus on the core algorithm of our linear solver, while \texttt{alpaka} takes care of porting and optimizing the code for the specific hardware.

\section{Methodology}
We select the algorithm for both the main solver and the preconditioner to develop a Poisson solver suitable for HPC frameworks. More specifically, we chose the algorithm to ensure compatibility with a matrix-free implementation and easy parallelization at the distributed memory level by dividing the computational domain into subdomains.

\subsection{Linear Solver \& Preconditioners}
The core algorithm used for solving the linear system in this work is the Bi-CGSTAB (we denote it also as BiCGS) method~\cite{vanDerVorst1992_BiCGstab}, which is an iterative Krylov solver introduced by Van der Vorst designed to solve large, sparse, and non-symmetric systems. The algorithm can be enhanced by applying preconditioning to accelerate convergence. A variant, known as Flexi-BiCGSTAB~\cite{vogel2007_flexiBiCGstab, Chen2016_flexiBiCGstab} introduces further flexibility in the preconditioning step, adapting it dynamically based on the current iteration. 

The preconditioned Bi-CGSTAB algorithm is outlined in Algorithm~\ref{algo_bicgstab}. In each iteration, the algorithm computes residual vectors, solves the preconditioned systems, and updates both the solution vector $\mathbf{x}$ and the residual $\mathbf{r}$, until the convergence criteria are met.
\begin{algorithm}[h!]
\caption{Preconditioned Bi-CGstab iteration to solve the linear system $\mathbf{A}\mathbf{x}=\mathbf{b}$}
\label{algo_bicgstab}
\begin{algorithmic}[1]
\State $\text{Compute } \mathbf{r_0}=\mathbf{b} - \mathbf{A}\mathbf{x_0} \text{ from initial guess } \mathbf{x_0}$
\State $\text{Choose } \mathbf{\Tilde{\mathbf{r}}} \text{ such that } \mathbf{\Tilde{\mathbf{r}}}^{T}\mathbf{r}_0\neq0, \text{ i.e. } \mathbf{\Tilde{\mathbf{r}}}=\mathbf{r}_0  $
\State $\mathbf{p}_0=\mathbf{r}_0$
\State $\rho_0 = \mathbf{\Tilde{\mathbf{r}}}^{T}\mathbf{r}_0$
\For{$i=1:\text{iterMax}$}
\State $\text{Solve } \mathbf{M}_i\mathbf{\hat{p}}_i=\mathbf{p}_{i-1}$
\State $\alpha_i = \dfrac{\rho_{i-1}}{\mathbf{\Tilde{\mathbf{r}}}^{T}\mathbf{A}\mathbf{\hat{p}}}_i$
\State $\mathbf{r}_i = \mathbf{r}_{i-1} - \alpha_i \mathbf{A}\mathbf{\hat{p}}_i$
\If{$||\mathbf{r_i}||<\text{tol}$}
\State $\mathbf{x}_i=\mathbf{x}_{i-1} + \alpha_i\mathbf{\hat{p}}_i$
\State $\text{break}$
\EndIf
\State $\text{Solve } \mathbf{M}_i\mathbf{\hat{r}}_i=\mathbf{r}_i$
\State $\mathbf{t} = \mathbf{A}\mathbf{\hat{r}}_i$
\State $\omega_i = \dfrac{\mathbf{t}^T\mathbf{r}_i }{\mathbf{t}^T\mathbf{t}}$
\State $\mathbf{r}_i = \mathbf{r}_i - \omega_i \mathbf{t}$
\State $\mathbf{x}_i=\mathbf{x}_{i-1} + \alpha_i\mathbf{\hat{p}}_i + \omega_i \mathbf{\hat{r}}_i$
\If{$||\mathbf{r_i}||<\text{tol}$}
\State $\text{break}$
\EndIf
\State $\rho_i = \mathbf{\Tilde{\mathbf{r}}}^{T}\mathbf{r}_i $
\State $\beta_i = \dfrac{\rho_i}{\rho_{i-1}}\dfrac{\alpha_i}{\omega_i}$
\State $\mathbf{p}_{i}= \mathbf{r}_i + \beta_i(\mathbf{p}_{i-1} -\omega_i\mathbf{A}\mathbf{\hat{p}}_i) $
\EndFor
\end{algorithmic}
\end{algorithm}

In the context of solving the linear system $\mathbf{A}\mathbf{x}=\mathbf{b}$, a main objective is to enhance the convergence rate of the solver improving the spectral properties of matrix $\mathbf{A}$. This can be achieved by finding a preconditioner matrix $\mathbf{M}$ such that $\mathbf{M}^{-1} \mathbf{A} \approx \mathbf{I}$. A good preconditioner should approximate well the inverse of the matrix $\mathbf{A}^{-1}$, as expressed by $\mathbf{M}^{-1} \approx \mathbf{A}^{-1}$, being at the same time inexpensive to compute.

Several preconditioning techniques have been tested in this work. The intrinsic flexibility of BiCGS allows us to use an inexact preconditioner, namely a preconditioner that, within certain limits, changes at each iteration. Thus, it is possible to exploit an iterative method, such as BiCGS itself, to obtain an approximate solution of the problem to use as a preconditioner \cite{Chen2016_flexiBiCGstab}. The preconditioner is $\mathbf{M}_i^{-1} \approx \mathbf{A}^{-1}$,
where the subscript $i$ denotes that the preconditioner matrix changes at each iteration. We denote this preconditioner as G(BiCGS) where G stands for global, and the resulting preconditioned solver as FBiCGS-G(BiCGS).

Despite being effective and easy to implement, this preconditioner has some drawbacks. As pointed out in Ref.~\cite{Chen2016_flexiBiCGstab}, it is not straightforward to set the exit conditions for the inner iterative solver since there is no precise theory about the minimum number of iterations or inner solver accuracy required to ensure the main solver convergence. Clearly, a theory about the optimal number of inner solver iterations (or accuracy) that guarantees the shortest time to solution is also not available. 
Additionally, G(BiCGS) is not communication-free, which is a significant drawback especially for GPU implementation. Indeed, in a distributed-memory implementation, grid points at the boundaries of each subdomain must be shared between the neighboring subdomains in order to build the approximate solution of the global problem $\mathbf{A}$. Moreover, iterative Krylov solvers, such as BiCGS, are not reduction-free. 

Since the parallel implementation of the solver relies on the division of the computational domain in subdomains, domain decomposition methods, like the additive Schwarz method (ASM) or the Restricted Additive Schwarz (RAS) method, arise as straightforward preconditioners~\cite{dolean2015_domainDecompositionSchwartz}. These techniques involve solving a local problem independently in each subdomain $s$, and then building the preconditioner matrix as the union of local solutions $\mathbf{A}_s^{-1}$. The power of Schwarz methods relies on the overlapping between subdomains, namely the global domain can be subdivided such that some grid points belong to multiple subdomains at the same time. For this reason, Schwarz methods are often employed in the solution of finite element problems~\cite{dolean2015_domainDecompositionSchwartz}.
Aiming for a communication-free preconditioner, we can build non-overlapping subdomains where no grid point belongs to more than one subdomain. In this case, ASM and RAS preconditioners reduce to the widely known and exploited block Jacobi (BJ) preconditioner. The BJ preconditioner is mathematically expressed as:
\begin{equation}
    \mathbf{M}^{-1}_{\text{BJ}} := \sum_{s}^{N_s} \mathbf{R}^T_s (\mathbf{R}_s\mathbf{A}\mathbf{R}^T_s)^{-1}\mathbf{R}_s
    \label{eq_blockJacobiPreconditioner}
\end{equation}
where $\mathbf{R}_s\in\Re^{N_{local}\times N_p}$ is the restriction matrix from the global number of unknowns ($N_p$) to the number of unknowns in each subdomain $s$ ($N_{local}$). Thus, $\mathbf{R}_s\mathbf{A}\mathbf{R}^T_s \in\Re^{N_{local}\times N_{local}} $ is the diagonal block of $\mathbf{A}$ corresponding to the unknowns in the subdomain $s$. The restriction operator applied to a vector and its preconditioned counterpart, denoted respectively as $\mathbf{p}$ and $\mathbf{\hat{p}}$, yields
$\mathbf{p}_s := \mathbf{R}_s \mathbf{p}$ and $\mathbf{\hat{p}}_s := \mathbf{R}_s \mathbf{\hat{p}}$, with $s=1,...,N_s$. Taking the expression for the preconditioner $\mathbf{M}_{\text{BJ}}^{-1}$ in Eq.~\ref{eq_blockJacobiPreconditioner}, we can write an explicit formula for the restricted preconditioned vector:
\begin{equation}
\begin{split}
    \mathbf{\hat{p}}_s & = \mathbf{R}_s \mathbf{M}^{-1}_{\text{BJ}} \mathbf{p} = \mathbf{R}_s \sum_{j}^{N_s} \mathbf{R}^T_j (\mathbf{R}_j\mathbf{A}\mathbf{R}^T_j)^{-1}\mathbf{R}_j \mathbf{p} = \\
    &=(\mathbf{R}_s\mathbf{A}\mathbf{R}^T_s)^{-1} \mathbf{R}_s \mathbf{p} = (\mathbf{R}_s\mathbf{A}\mathbf{R}^T_s)^{-1} \mathbf{p}_s, \quad s=1,...,N_s
\end{split}
    \label{eq_restrictedPreconditionedVector}
\end{equation}
where we have exploited the non-overlapping domain decomposition, namely $\mathbf{R}_s\mathbf{R}^T_j = \mathbf{0}$ if $j\neq s$. Eq.~\ref{eq_restrictedPreconditionedVector} gives an important result: in case of no overlapping, no global information is needed to compute the preconditioned vector. Indeed, the preconditioner can be applied independently to the restricted vector in each subdomain requiring only the local information, making it inherently communication-free.
Since we aim for a matrix-free solver, we avoid building explicitly the preconditioner matrices $(\mathbf{R}_s\mathbf{A}\mathbf{R}^T_s)^{-1}$, directly computing the action of the preconditioner on a restricted vector as a solution of the linear system
\begin{equation}
    (\mathbf{R}_s\mathbf{A}\mathbf{R}^T_s) \mathbf{\hat{p}}_s = \mathbf{p}_s \quad \text{with }s=1,...,N_s .
    \label{eq_systemBJpreconditioner}
\end{equation}
Thus, the application of the BJ preconditioner results in solving locally in each subdomain $s$ a linear system that corresponds to a restricted version of the global problem.

The linear system in Eq.~\ref{eq_systemBJpreconditioner} is typically solved with exact methods (e.g. LU factorization), but this approach is not suitable for a matrix-free solver. Moreover, classical parallel CPU solvers divide the domain into many small subdomains, whereas GPU solvers use fewer, larger ones, making exact methods impractical due to their high computational cost. Once again, Krylov iterations emerge as promising options for computing the preconditioner. A straightforward approach is to utilize the already available BiCGS method, leading to a preconditioner we denote as BJ(BiCGS). The corresponding solver is referred to as FBiCGS-BJ(BiCGS). In this case, there are no communication drawbacks and no global reductions between subdomains due to the locality of the information. However, issues related to the optimal number of iterations, or equivalently to the required accuracy of the solution, persist, and it is not easy to find the best conditions to exit the inner iteration cycle. Moreover, even though there is no global reduction between subdomains, BiCGS remains a non-reduction-free solver since it requires scalar products, and this might raise numerical issues due to floating-point arithmetic.

The two preconditioning techniques described so far can be applied to solve any linear system $\mathbf{A}\mathbf{x}=\mathbf{b}$ since no specific knowledge about the matrix operator $\mathbf{A}$, besides the matrix itself, is needed. However, since we are building a solver specifically for the Poisson equation, we can trade the solver generality with higher performance.
In particular, provided that the eigenvalues of the matrix $\mathbf{A}$ are known, it is possible to leverage Chebyshev polynomials to build an iterative solver~\cite{saad2003_iterativeSparseLinearSystems,Gutknecht2002_chebyshev}. The resulting iteration, called Chebyshev iteration (CI), can be used as the main solver to compute the general problem solution. However, its convergence rate is known to be slower compared to iterative Krylov methods. Therefore, it is more often employed as a preconditioner~\cite{saad2003_iterativeSparseLinearSystems,bergamaschi2023_polynomialPreconditioner}. The Chebyshev iteration can be applied to symmetric and non-symmetric matrices whose eigenvalues are confined to an elliptic domain that does not include the origin. Assuming real eigenvalues in the domain $\lambda\in[\alpha,\beta]$, and defining the following three quantities:
\begin{equation}
    \theta=\dfrac{\beta + \alpha}{2}, \quad \delta=\dfrac{\beta - \alpha}{2}, \quad \sigma=\dfrac{\theta}{\delta},
    \label{eq_chebyshevQuantities}
\end{equation}
we have the iterative algorithm reported in Algorithm~\ref{algo_chebyshev}.
\begin{algorithm}
\caption{Chebyshev iteration to solve the system $\mathbf{A}\mathbf{x}=\mathbf{b}$, where $\theta$, $\delta$, $\sigma$ are defined in Eq. \ref{eq_chebyshevQuantities}}
\label{algo_chebyshev}
\begin{algorithmic}[1]
\State $\rho_{old}={1}/{\sigma}$
\State $\rho_{cur}={1}/({2\sigma - \rho_{old}})$
\State $\mathbf{z}={\mathbf{b}}/{\theta} \quad (\text{if iterMax}=0\text{, exit with } \mathbf{x}=\mathbf{z} ) $ 
\State $\mathbf{y} = 2\dfrac{\rho_{cur}}{\delta}\left(2\mathbf{b} - \dfrac{\mathbf{A}\mathbf{b}}{\theta} \right) (\text{if iterMax}=1\text{, exit with } \mathbf{x}=\mathbf{y})$ 
\For{$i=2:\text{iterMax}$}
\State $\rho_{old} = \rho_{cur}$
\State $\rho_{cur}={1}/({2\sigma - \rho_{old}})$
\State $\mathbf{w} =\rho_{cur}\left( 2\sigma\mathbf{y} + \dfrac{2}{\delta}(\mathbf{b} - \mathbf{A\mathbf{y}}) - \rho_{old}\mathbf{z} \right) $
\State $\mathbf{z}=\mathbf{y}$
\State $\mathbf{y} = \mathbf{w}$
\EndFor
\State $\mathbf{x} = \mathbf{w} $
\end{algorithmic}
\end{algorithm}

This iteration fits perfectly our problem since only the smaller and the larger eigenvalues of the system are needed, which can be easily computed in the case of the Poisson matrix as shown in Eqs.~\ref{eq_lambdaMin} and~\ref{eq_lambdaMax}. Moreover, the Chebyshev algorithm is inherently reduction-free, which makes it a fixed preconditioner and allows interesting modifications suitable for a parallel framework, as will be explained later.

CI can be used to solve the small local problem, arising from the BJ preconditioner, Eq.~\ref{eq_systemBJpreconditioner}, leading to a solver that we denote as BiCGS-BJ(CI). Since we are solving a preconditioner problem, an exact solution of Eq.~\ref{eq_systemBJpreconditioner} is not needed, and we can stop the CI after only few iterations. Note that, differently from the BJ(BiCGS) preconditioner, BJ(CI) is a fixed preconditioner independent from the number of CI iterations employed since CI is a reduction-free solver.

As previously described for FBiCGS-G(BiCGS), we can also build the preconditioner as an approximate solution of the global problem, $\mathbf{M}^{-1}\approx\mathbf{A}^{-1}$. The CI can be exploited to obtain this approximate solution, leading to a preconditioner that we denote as G(CI) and the corresponding solver BiCGS-G(CI). Also in this case, since CI is reduction-free, G(CI) is a fixed preconditioner.

While BJ(CI) is a communication-free preconditioner, G(CI) is not due to neighboring subdomains sharing grid point values at boundaries. However, since the global approximate solution is employed as a preconditioner, some information loss in its computation may be acceptable. Thus, thanks to CI being reduction-free, we can obtain a new preconditioner simply by avoiding the communication between subdomains in G(CI). This new preconditioner is clearly both communication-free and reduction-free. We denote it as GNoComm(CI), where NoComm stands for no communication, and we call the corresponding solver BiCGS-GNoComm(CI). GNoComm(CI) is equivalent to the BJ(CI) preconditioner in which the parameters in Eq.~\ref{eq_chebyshevQuantities} are computed with the eigenvalues of the global matrix $\mathbf{A}$ instead of the eigenvalues of the local system $(\mathbf{R}_s\mathbf{A}\mathbf{R}^T_s)$.
There is no available theory on the optimal number of CI preconditioner iterations that minimizes the overall solver time to solution. However, we can derive an upper bound for the number of iterations in the case of GNoComm(CI). Since there is no communication between neighboring subdomains, the ghost points are not updated and their values can be considered "incorrect". The stencil operator applied to boundary points takes the incorrect ghost cell values to perform computations, leading to a partially incorrect update, and at each following iteration the stencil propagates this error towards the center of the subdomain. Given a $N_s\times N_s\times N_s$ subdomain, after $N_s/2$ iterations, the error has propagated to all points, and from here on, all point updates will be partially incorrect. Thus, we can set $N_s/2$ as the upper bound for the number of iterations in GNoComm(CI).

Table \ref{tab_testedSolvers} summarizes the solvers tested in this work.

\begin{table}[h!]
\centering
\caption{Summary of the tested solvers characteristics}
\begin{tabular}{|c|c|c|c|}
\hline
\multirow{2}{*}{\textbf{Solver}}   & \textbf{Fixed} & \textbf{Comm-free} & \textbf{Reduction-free} \\
&  \textbf{prec.} & \textbf{prec.} & \textbf{prec.} \\ \hline
BiCGS &   - & - & -  \\ \hline
FBiCGS-G(BiCGS) &   {\color{red} \xmark} & {\color{red} \xmark } & {\color{red} \xmark } \\ \hline
FBiCGS-BJ(BiCGS) &   {\color{red} \xmark }& {\color{Green} \checkmark } & {\color{red} \xmark }\\ \hline
BiCGS-BJ(CI) &   {\color{Green} \checkmark }  & {\color{Green} \checkmark }  &  {\color{Green} \checkmark }  \\ \hline
BiCGS-G(CI) &   {\color{Green} \checkmark } & {\color{red} \xmark } &  {\color{Green} \checkmark }  \\ \hline
BiCGS-GNoComm(CI) &  {\color{Green} \checkmark } & {\color{Green} \checkmark }  &  {\color{Green} \checkmark }  \\ \hline
\end{tabular}
\label{tab_testedSolvers}
\end{table}

\subsection{Overall Algorithms}
The Bi-CGSTAB algorithm (Algorithm~\ref{algo_bicgstab}) involves different operations, including the computation of the residuals, application of the preconditioner, matrix-vector multiplications, and updates of the solution vector. It also requires communication between different subdomains via MPI to share data, such as the halo cells, and perform the necessary reductions across processes. To improve efficiency and memory utilization, we implement a matrix-free solver, where the matrices are not explicitly stored, and matrix-vector multiplications are performed applying a stencil to the data.
Rearranging and merging the operations in the original algorithm enhances both temporal and spatial data locality, leading to the final implementation outlined in the pseudocode in Algorithm~\ref{algo_bicgstabImplementation}. The names of the functions that implement each step are shown on the right. 

\begin{algorithm}
\caption{Bi-CGstab implementation pseudocode}
\label{algo_bicgstabImplementation}
\begin{algorithmic}[1]
\State $\text{Compute } \mathbf{r_0}=\mathbf{b} - \mathbf{A}\mathbf{x_0} \text{ from initial guess } \mathbf{x_0}$
\State $\text{Choose } \mathbf{\Tilde{\mathbf{r}}} \text{ such that } \mathbf{\Tilde{\mathbf{r}}}^{T}\mathbf{r}_0\neq0, \text{ i.e. } \mathbf{\Tilde{\mathbf{r}}}=\mathbf{r}_0  $
\State $\mathbf{p}_0=\mathbf{r}_0$
\State $\rho_0 = \mathbf{\Tilde{\mathbf{r}}}^{T}\mathbf{r}_0$
\For{$i=1:\text{iterMax}$}
\State Solve $\mathbf{M}_i\mathbf{\hat{p}}_i=\mathbf{p}_{i-1}$ \Comment{\colorbox{highlight8}{ Preconditioner}}
\State MPI communicate halo-points $\mathbf{\hat{p}}_i^s$ \Comment{\colorbox{highlight4}{MPI1}}
\State Set Neumann BCs to $\mathbf{\hat{p}}_i^s$ \Comment{\colorbox{highlight7}{KernelNeumannBCs}}
\State $\mathbf{w}^s=\mathbf{A}\mathbf{\hat{p}}_i^s$ \Comment{\colorbox{highlight2}{KernelBiCGS1}}
\State Local  $p_{sum}^s=\mathbf{\Tilde{\mathbf{r}}}^{Ts}\mathbf{w}^s$ \Comment{\colorbox{highlight2}{KernelBiCGS1}}
\State MPI\_Allreduce $p_{sum}^{\text{tot}}=\sum_s p_{sum}^{s} $ \Comment{\colorbox{highlight9}{MPI2}}
\State $\alpha_i = \rho_{i-1}/p_{sum}^{\text{tot}}$
\State $\mathbf{r}_i^s = \mathbf{r}_{i-1}^s - \alpha_i \mathbf{w}^s$ \Comment{\colorbox{highlight10}{KernelBiCGS2}}
\State Solve $\mathbf{M}_i\mathbf{\hat{r}}_i=\mathbf{r}_i$ \Comment{\colorbox{highlight8}{ Preconditioner}}
\State MPI communicate halo-points $\mathbf{\hat{r}}_i^s$ \Comment{\colorbox{highlight4}{MPI3}}
\State Set Neumann BCs to $\mathbf{\hat{r}}_i^s$ \Comment{\colorbox{highlight7}{KernelNeumannBCs}}
\State $\mathbf{t}^s = \mathbf{A}\mathbf{\hat{r}}_i^s$ \Comment{\colorbox{highlight3}{KernelBiCGS3}}
\State Local  $p1_{sum}^s=\mathbf{{\mathbf{t}}}^{Ts}\mathbf{r}_i^s$ \Comment{\colorbox{highlight3}{KernelBiCGS3}}
\State Local  $p2_{sum}^s=\mathbf{{\mathbf{t}}}^{Ts}\mathbf{t}^s$ \Comment{\colorbox{highlight3}{KernelBiCGS3}}
\State MPI\_Allreduce $p1_{sum}^{\text{tot}}=\sum_s p1_{sum}^{s} $ \Comment{\colorbox{highlight9}{MPI4}}
\State MPI\_Allreduce $p2_{sum}^{\text{tot}}=\sum_s p2_{sum}^{s} $\Comment{\colorbox{highlight9}{MPI4}}
\State $\omega_i = p1_{sum}^{\text{tot}}/p2_{sum}^{\text{tot}}$
\State $\mathbf{x}_i^s=\mathbf{x}_{i-1}^s + \alpha_i\mathbf{\hat{p}}_i^s + \omega_i \mathbf{\hat{r}}_i^s$ \Comment{\colorbox{highlight6}{KernelBiCGS4}}
\State $\mathbf{r}_i^s = \mathbf{r}_i^s - \omega_i \mathbf{t}^s$ \Comment{\colorbox{highlight1}{KernelBiCGS5}}
\State Local  $p1_{sum}^s=\mathbf{{\mathbf{r_0}}}^{Ts}\mathbf{r}_i^s$ \Comment{\colorbox{highlight1}{KernelBiCGS5}}
\State Local  $p2_{sum}^s=\mathbf{{\mathbf{r_i}}}^{Ts}\mathbf{r}_i^s$ \Comment{\colorbox{highlight1}{KernelBiCGS5}}
\State MPI\_Allreduce $p1_{sum}^{\text{tot}}=\sum_s p1_{sum}^{s} $ \Comment{\colorbox{highlight9}{MPI5}}
\State MPI\_Allreduce $p2_{sum}^{\text{tot}}=\sum_s p2_{sum}^{s} $ \Comment{\colorbox{highlight9}{MPI5}}
\If{$p2_{sum}^{\text{tot}}<\text{tol}$}
\State $\text{break}$
\EndIf
\State $\rho_i = p1_{sum}^{\text{tot}} $
\State $\beta_i = ({\rho_i}\alpha_i)/({\rho_{i-1}}{\omega_i})$
\State $\mathbf{p}_{i}^s= \mathbf{r}_i^s + \beta_i(\mathbf{p}_{i-1}^s -\omega_i\mathbf{w}^s) $ \Comment{\colorbox{highlight5}{KernelBiCGS6}}
\EndFor
\end{algorithmic}
\end{algorithm}

The pseudocode in Algorithm~\ref{algo_chebyshevImplementation} outlines the implementation of the Chebyshev iteration. In the algorithm, the quantities $\theta$, $\delta$, and $\sigma$ are defined in Eq.~\ref{eq_chebyshevQuantities} and play a crucial role in the calculation of the iterative steps. 
In both Algorithms~\ref{algo_bicgstabImplementation} and~\ref{algo_chebyshevImplementation} the superscripts $s$ indicate the restriction of vectors in the subdomain $s$.

\subsection{\texttt{alpaka} Implementation}
The solver has been developed with efficiency and cross-platform portability as primary goals. In order to exploit HPC machines in the most effective way, the solver, written in C++, has been developed to natively support distributed memory architectures and GPU acceleration. The former goal is achieved thanks to the MPI, 
while GPU acceleration is achieved via the \texttt{alpaka} portability framework~\cite{alpaka2015,alpaka2016,alpaka2017}.
The code snippet reported in Listing~\ref{lst_alpaka} shows how a kernel is defined and launched in \texttt{alpaka}. The actual hardware is specified at line $1$ (AMD GPUs with HIP in this case), while everything else in the code is back-end independent.

\begin{lstlisting}[style=cppstyle, caption=Example alpaka kernel definition and launching., label=lst_alpaka]
// Define the GPU accelerator type
using Acc = alpaka::AccGpuHipRt<Dim, Idx>; 
// Get the GPU device
alpaka::Dev<alpaka::Platform<Acc>> devAcc = 
    alpaka::getDevByIdx(alpaka::Platform<Acc>{}, 0);
// Create a command queue for the solver 
alpaka::Queue<Acc, alpaka::Blocking> queueSolver;
// Define the computational domain (grid size)
alpaka::Vec<DIM, Idx> extent = {N_x, N_y, N_z};
// Instantiate the BiCGStab solver kernel
BiCGS1Kernel<DIM, T_data> biCGS1Kernel;
// Compute the valid work division for the kernel execution
alpaka::WorkDivMembers<DIM, Idx> workDivKernel1 = alpaka::getValidWorkDiv(
    extent,    
    devAcc,    // Target device
    biCGS1Kernel, // Kernel function
    /* kernel parameters */);
// Launch the kernel asynchronously on the GPU
alpaka::exec<Acc>(
    queueSolver,     // Execution queue
    workDivKernel1,  // Work division
    biCGS1Kernel,    // Kernel function
    ... /* kernel parameters */);
\end{lstlisting}

\begin{algorithm}
\caption{Chebyshev iteration implementation pseudocode.}
\label{algo_chebyshevImplementation}
\begin{algorithmic}[1]
\State $\rho_{old}={1}/{\sigma}$
\State $\rho_{cur}={1}/{2\sigma - \rho_{old}}$
\State MPI communicate halo-points $\mathbf{b}$ \Comment{\colorbox{highlight4}{MPI1}}
\State Set Neumann BCs to $\mathbf{b}$ \Comment{\colorbox{highlight7}{KernelNeumannBCs}}
\State $\mathbf{z}^s={\mathbf{b}^s}/{\theta}$ \Comment{\colorbox{highlight3}{KernelCI1}}
\State $\mathbf{y}^s = 2\dfrac{\rho_{cur}}{\delta}\left(2\mathbf{b}^s - \dfrac{\mathbf{A}\mathbf{b}^s}{\theta} \right)$ \Comment{\colorbox{highlight3}{KernelCI1}}
\For{$i=2:\text{iterMax}$}
\State $\rho_{old} = \rho_{cur}$
\State $\rho_{cur}={1}/({2\sigma - \rho_{old}})$
\State MPI communicate halo-points $\mathbf{y}$ \Comment{\colorbox{highlight4}{MPI2}}
\State Set Neumann BCs to $\mathbf{y}$ \Comment{\colorbox{highlight7}{KernelNeumannBCs}}
\State $\mathbf{w}^s =\rho_{cur}\left( 2\sigma\mathbf{y}^s + \dfrac{2}{\delta}(\mathbf{b}^s - \mathbf{A}\mathbf{y}^s) - \rho_{old}\mathbf{z}^s \right) $ \Comment{\colorbox{highlight1}{KernelCI2}}
\State $\mathbf{z}^s=\mathbf{y}^s$  \Comment{\colorbox{highlight9}{Swap pointers}}
\State $\mathbf{y}^s = \mathbf{w}^s$ \Comment{\colorbox{highlight9}{Swap pointers}}
\EndFor
\State $\mathbf{x} = \mathbf{w} $ \Comment{\colorbox{highlight5}{KernelCI3}}
\end{algorithmic}
\end{algorithm}

\begin{figure}[h!]
    \centering
    \includegraphics[width=0.99\linewidth]{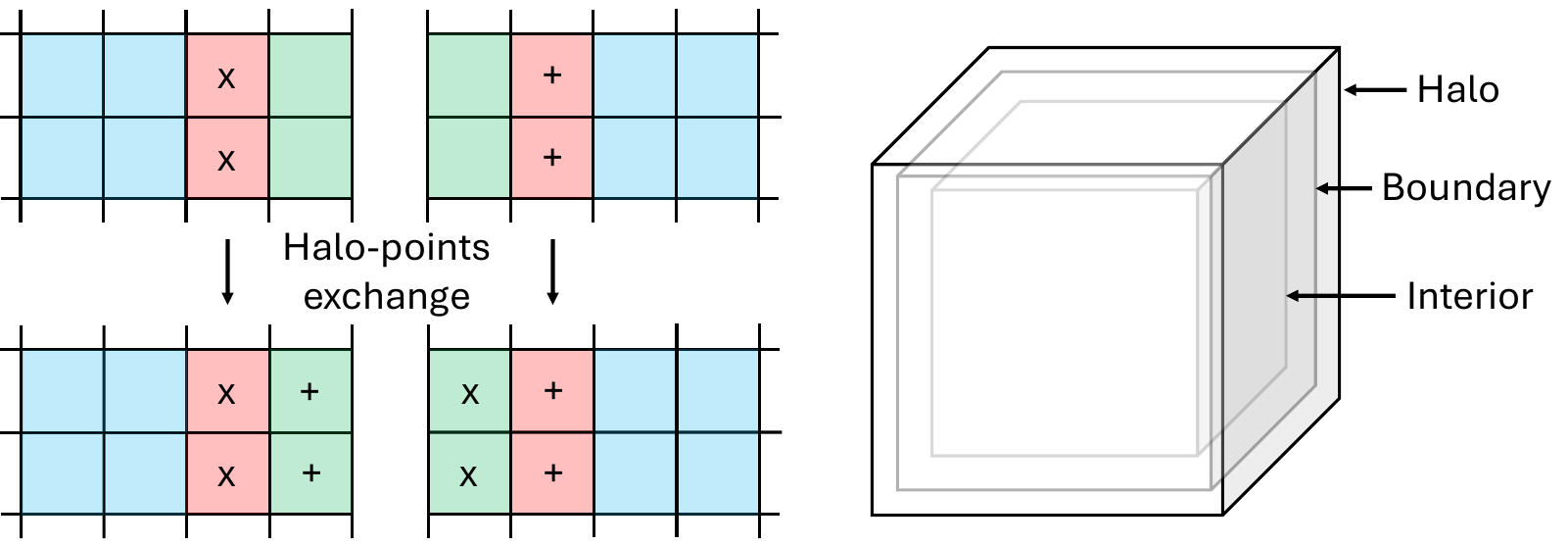}
    \caption{Halo-points communication between neighboring subdomains.}
    \label{fig_haloPointsExchange}
\end{figure}

Parallelism is achieved by dividing the global domain into subdomains of equal dimensions, each assigned to a different MPI process. The number of subdomains in each direction ($N_s^x$, $N_s^y$, $N_s^z$) can be freely set by the user, with the only requirement that $N_s^x \times N_s^y \times N_s^z = N_{\text{MPI}}^{\text{tot}} $, where $N_{\text{MPI}}^{\text{tot}}$ is the total number of MPI ranks.
Each MPI rank stores the local data of its subdomain, plus data that correspond to the border grid points of the neighboring subdomains, which constitute the so-called grid halo (ghost) points. Thus, the local grid size in each subdomain in each direction $i$ is equal to $N_{\text{local}}^i = N_{\text{global}}^i/N_s^i + 2N_{\text{halo}}^i$. In the case of the Poisson equation discretized with a 
second-order centered finite difference scheme, $N_{\text{halo}}=1$ for each direction. Fig.~\ref{fig_haloPointsExchange} shows the halo exchange between neighboring processes.

The solver relies on a workhorse class called \verb|blockGrid| that stores all the information about the global domain and the local subdomain, such as the number of grid points and the subdomain location in the grid. This class is particularly useful for implementing communications between MPI ranks.

The solver leverages the built-in \verb|MPI_Datatype| class so that grid points in the border region of each subdomain are aggregated in a single data structure that can be sent and received with a single MPI call. Communications are managed independently for each direction and side of the subdomain. For instance, in the case of a 3D simulation, a subdomain neighboring with six other subdomains will share its border grid points through six MPI\_send calls and will fill in its halo points through six MPI\_recv calls. Asynchronous communication calls (\verb|MPI_Isend| and \verb|MPI_Irecv|) are implemented, and synchronization is ensured at the end of the communication stage with the \verb|MPI_Waitall| function.

Data communication, notoriously an issue in HPC applications, is even more problematic when dealing with GPU-accelerated codes, where, besides communication between different MPI processes, the data movement between GPU and CPU in each process must be handled.
In the solver, GPUs perform the majority of computations. Therefore, the data, namely the arrays that store vectors, are offloaded to the accelerators at the beginning of the simulation and copied back to the CPUs only once when the solution is obtained. Only a few operations are performed by the CPU. In the BiCGS loop (Algorithm~\ref{algo_bicgstab}), CPUs compute the coefficients $\alpha_i$, $\beta_i$ and $\omega_i$. However, the scalar products are evaluated on the GPUs. In the CI loop (Algorithm~\ref{algo_chebyshev}), only $\rho_{cur}$ is computed on the CPU. In order to share between different MPI ranks data that is stored in the GPU memory without explicitly copying it to the CPU, GPU-aware MPI and GPU-direct communication are exploited.

\section{Experimental Setup}
The problem under consideration is governed by the following Poisson equation:
\begin{equation}
    -\Delta\phi (x,y,z) = \sin{(x)} + \cos{(y)} + 3\sin{(z)} - 2yz + 2
\end{equation}
defined on the domain $x\in[3,28.5]$, $y\in[2.5,28]$, $ z\in[10,35.5]$. The mesh resolution is $\Delta x = \Delta y = \Delta z = 0.1$, resulting in a uniform Cartesian grid $256\times256\times256$. Dirichlet boundary conditions are applied on $\delta_{x^-}$, $\delta_{y^+}$, $\delta_{z^+}$, while Neumann boundary conditions are applied on $\delta_{x^+}$, $\delta_{y^-}$, $\delta_{z^-}$. We always normalize the right-hand side, making all tolerances relative tolerances. The stopping criteria for the main solver iteration is $1\times10^{-10}$. The preconditioner stopping criteria for the various methods are as follows:
\begin{itemize}
\item For FBiCGS-G(BiCGStab), the stopping tolerance is set to $1\times10^{-2}$, with a maximum of 500 iterations.
\item For FBiCGS-BJ(BiCGStab), the stopping tolerance is set to $1\times10^{-6}$, with a maximum of 500 iterations.
\item For CI-based preconditioners, a fixed number of 24 iterations is used derived from the theory.
\end{itemize}
Bergamaschi \textit{et Al.}~\cite{bergamaschi2023_polynomialPreconditioner} has shown that computing the CI with rescaled eigenvalues $\lambda_{\min}'$ and $\lambda_{\max}'$, that are slightly inside the interval $[\lambda_{\min}, \lambda_{\max}]$,     can greatly reduce the number of main solver iterations. Thus, for BiCGS-G(CI) and BiCGS-GNoComm(CI), we rescale the maximum and minimum eigenvalue of the matrix by $(1-1\times10^{-4})$ and 100 respectively.

For the performance evaluation, we use two accelerated EuroHPC supercomputers: LUMI-G (equipped with AMD MI250X GPUs) and MareNostrum5 (featuring NVIDIA Hopper H100 GPUs). The solver is also tested on CPUs on the non-accelerated partition of LUMI. These systems are used to demonstrate the performance of our code across different hardware architectures.
The calculations are performed with the following configurations:
\begin{itemize}
    \item 64 AMD MI250X GPU Compute Devices (GCDs), each managed by one MPI process, distributed across 8 nodes on LUMI-G. 
    \item 64 NVIDIA Hopper H100 GPUs, each managed by one MPI process, distributed across 16 nodes on MareNostrum5-G. 
    \item 1,024 OpenMP threads, managed by 64 MPI processes, distributed across 8 nodes on LUMI-C EuroHPC. 
\end{itemize}

More specifically, all solvers are tested on LUMI-G, then, we conduct a deeper analysis on the most efficient one (BiCGS-GNoComm(CI)) evaluating its performances across different CPU and GPU architectures.

In order to better characterize BiCGS-GNoComm(CI), decoupling computation from communication, we also test the code on a smaller mesh ($64\times64\times64$) using one single MPI process, thus running with only one device for the GPU implementations, and with 128 OMP threads for the CPU one. Note that in this case the domain is not subdivided in subdomains. Therefore, BJ preconditioners would coincide to the global ones. We fix the number of preconditioner iterations to 24 and we rescale the maximum and minimum eigenvalue of the matrix by  $(1-1\times10^{-4})$ and 10 respectively.

Eventually, we test the BiCGS-GNoComm(CI) solver scalability against a larger problem ($1024\times 1024\times 1024$ mesh) on LUMI-G.
\section{Results}

\subsection{Convergence, Solver \& Preconditioners Performance}
We evaluate the effectiveness of the preconditioning techniques by analyzing the convergence behavior of the solvers and the total time to solution for the problem with a $256 \times 256 \times 256$ mesh. All solvers are tested on the LUMI-G supercomputer using 64 GCDs and 64 MPI processes.

\begin{figure}[h!]
    \centering  
    \includegraphics[width=0.98\linewidth]{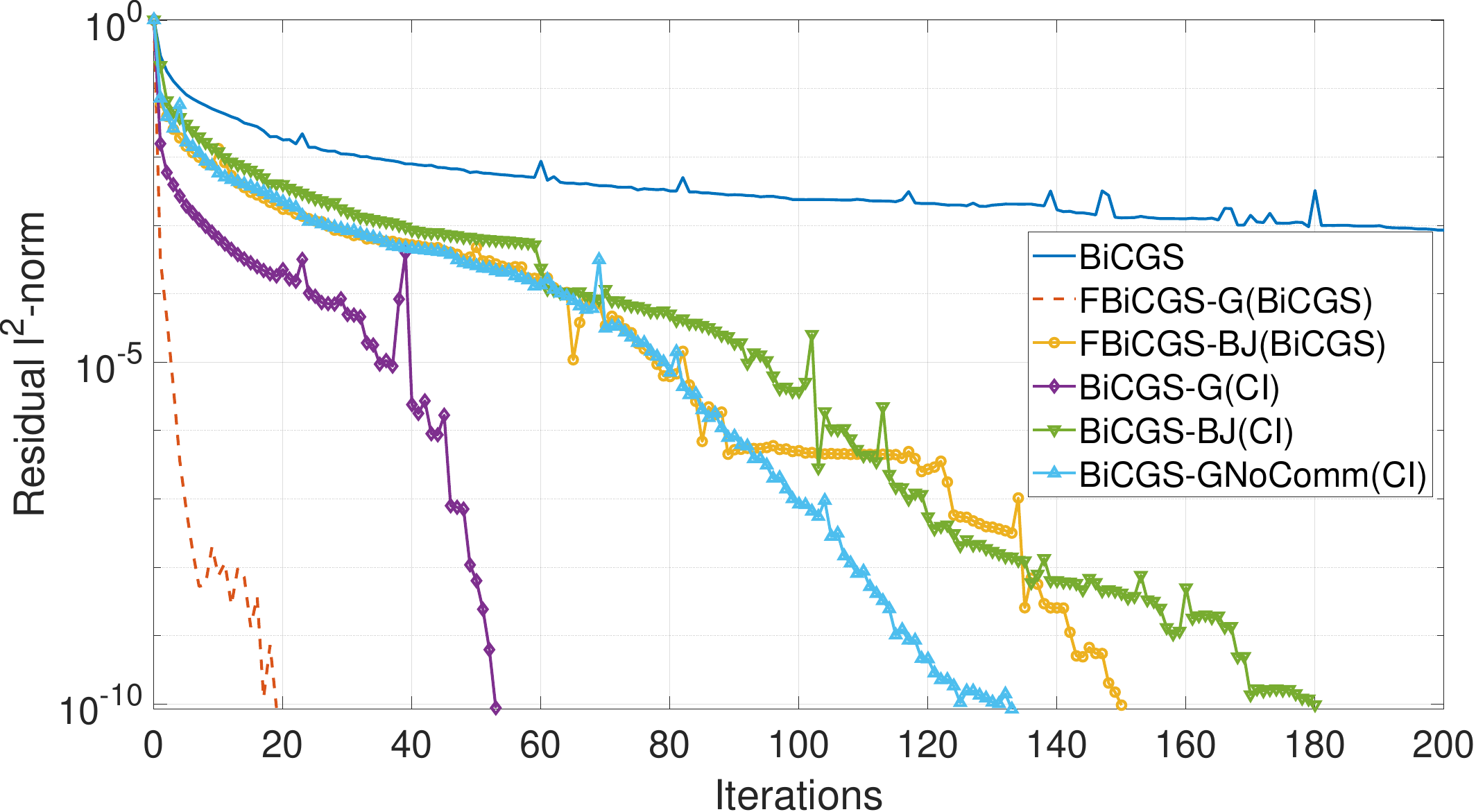}
    \caption{Residual norm at each iteration for the different tested solvers for a $256\times256\times256$ mesh. Results obtained running the solvers on 64 GCDs on LUMI-G with 64 MPI processes.}
    \label{fig_reisdualHistoryMesh256_MPI64_solvers}
\end{figure}

Fig.~\ref{fig_reisdualHistoryMesh256_MPI64_solvers} shows the residual norm as a function of the solver iteration, while in Table~\ref{tab_testedSolvers_results} the performance results of the different solvers, averaged over five runs, are detailed. In particular, the table reports the average number of outer iterations, the average number of preconditioner iterations per outer cycle, and the average total time to solution (TTS) for each solver tested. The results demonstrate that all the preconditioning methods significantly improve the solver's convergence rate, reducing the number of iterations required to achieve a $1\times10^{-10}$ error from 1600 to less than 200. While global preconditioners yield a solution in fewer main solver iterations (approximately 15 iterations for G(BiCGS)), the communication-free BiCGS-GNoComm(CI) solver achieves the lowest overall time to solution ($0.77$ s), making it the optimal algorithm. Specifically, BiCGS-GNoComm(CI) reaches the solution 6.5 times faster than the un-preconditioned solver and 50 times faster than the BiCGS-G(BiCGS) algorithm. Note that, the nondeterminism in floating point arithmetic, especially when reductions are involved, combined with the nonmonotonic behavior of the residual of the BiCGS algorithm, yields variance in the number of solver iterations required to reach convergence across different runs.

\begin{table}[h!]
\centering
\caption{Results for the test problem obtained on LUMI-G with the different solvers for a $256\times256\times256$ mesh and $64$ GCDs. }
\begin{tabular}{|c|c|c|c|}
\hline
\multirow{2}{*}{\textbf{Solver}} &  \textbf{Outer} & \textbf{Preconditioner} & \multirow{2}{*}{\textbf{TTS [s]}}  \\
& \textbf{it.} & \textbf{it. / outer it.} & \\ \hline
BiCGS & $1543\pm245$ & - & $5.0\pm0.8$ \\ \hline
FBiCGS-G(BiCGS) & $13\pm3$ & $950\pm10$ & $38\pm8$ \\ \hline
FBiCGS-BJ(BiCGS) & $125\pm12$ & $370\pm2$ & $35\pm3$ \\ \hline
BiCGS-BJ(CI) & $172\pm20$ & $48$ & $1.0\pm0.1$ \\ \hline
BiCGS-G(CI) & $50\pm2$ & $48$ & $3.3\pm0.1$\\ \hline
BiCGS-GNoComm(CI)& $140\pm12$ & $48$ & $0.77\pm0.06$ \\ \hline
\end{tabular}
\label{tab_testedSolvers_results}
\end{table}

Once BiCGS-GNoComm(CI) is identified as the optimal solver, we aim to characterize its convergence behavior across various hardware architectures, both at the node and multi-node level. Specifically, we test the BiCGS-GNoComm(CI) solver on LUMI-C, LUMI-G, and MareNostrum5.

\begin{figure}[!h]
    \centering
    \includegraphics[width=0.98\linewidth]{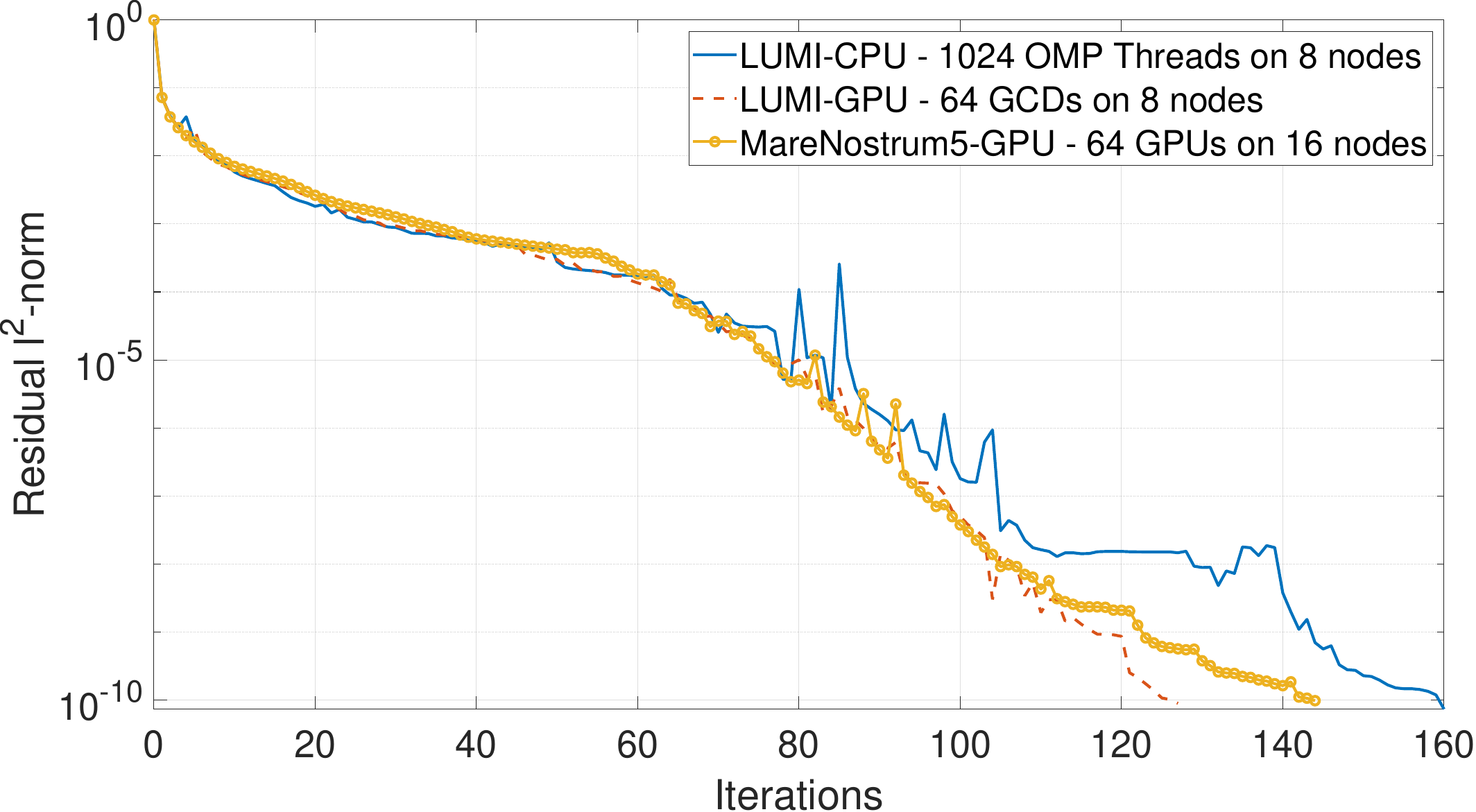}
    \caption{Residual norm at each iteration for BiCGS-GNoComm(CI) tested on different hardware architectures. $256\times256\times256$ mesh, 64 MPI processes.}
\label{fig_residualHistoryLUMIC_LUMIG_MN5G_Mesh256_MPI64}
\end{figure}

\begin{figure}[!h]
    \centering
    \includegraphics[width=0.98\linewidth]{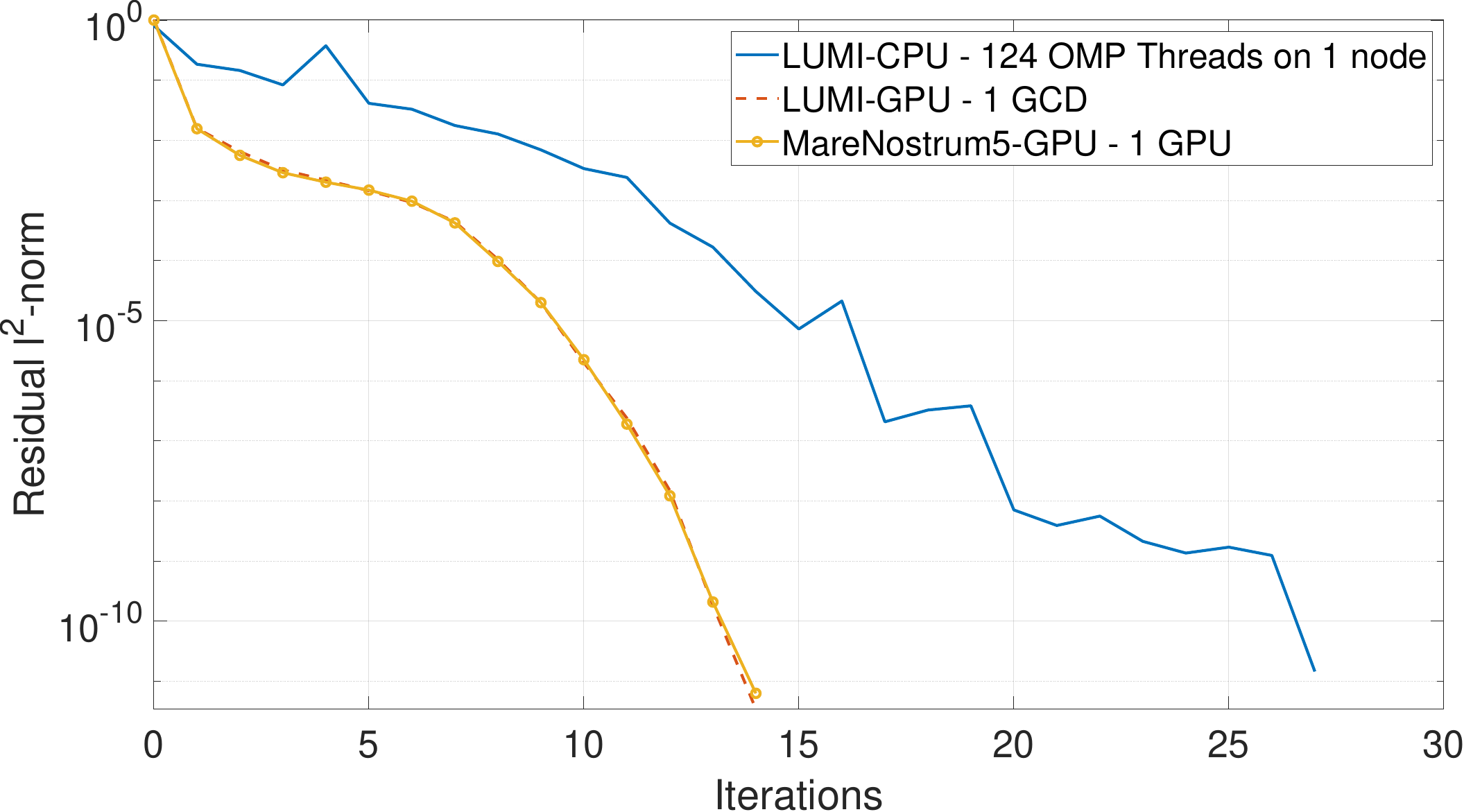}
    \caption{Residual norm at each iteration for BiCGS-GNoComm(CI) tested on different hardware architectures. $64\times64\times64$ mesh and one MPI process.}    \label{fig_residualHistoryLUMIC_LUMIG_MN5G_Mesh64_MPI1}
\end{figure}

Fig.~\ref{fig_residualHistoryLUMIC_LUMIG_MN5G_Mesh256_MPI64} shows the residual norm at each iteration for the test case with a $256 \times 256 \times 256$ mesh, run with 64 MPI processes. The convergence behavior in the case of the smaller problem, ($64\times64\times64$ mesh) run with a single MPI process, is reported in Fig.~\ref{fig_residualHistoryLUMIC_LUMIG_MN5G_Mesh64_MPI1}. In both experiments, the solver maintains the same convergence rate across AMD MI250X and NVIDIA Hopper H100 GPUs, while its performance is slightly lower when running on CPUs with OpenMP as the back-end. This is particularly noticeable in the single-process simulation where the CPU implementation requires 27 iterations to reach convergence, compared to 14 iterations with the AMD and NVIDIA back-ends. 

\observation{\textit{All the tested preconditioners enhance the BiCGSTAB convergence rate, reducing the required solver iterations from 1600 to fewer than 200. Global preconditioners are the most effective, however, the lowest time to solution is achieved with the communication-free GNoComm(CI) preconditioner. The BiCGS-GNoComm(CI) solver reaches the solution $6.5\times$ faster than the un-preconditioned BiCGSTAB and $50\times$ faster than FBiCGS-G(BiCGS). The convergence properties of BiCGS-GNoComm(CI) remain consistent across both NVIDIA Hopper H100 and AMD MI250X GPUs, whether executed with one or multiple MPI processes. Performance slightly degrades for the CPU back-end, especially in the single-node test, where it requires twice as many iterations as the GPU.}}

\subsection{Performance Analysis and Scalability}
\begin{figure}[h!]
    \centering
    \includegraphics[width=0.98\linewidth]{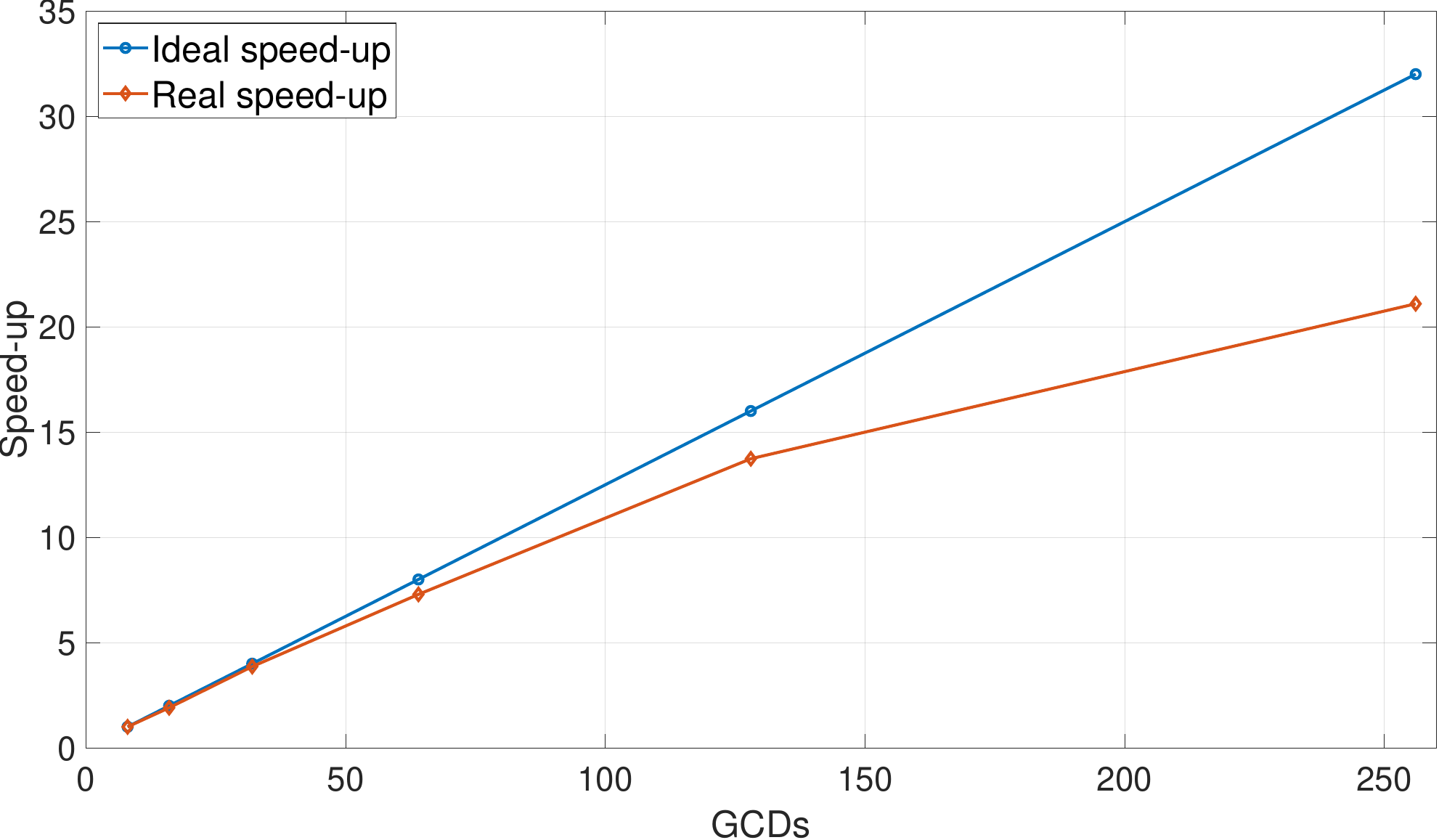}
    \caption{Strong scaling test performed on LUMI-G with a $1024\times 1024\times1024$ mesh. The reference is the time to solution obtained with 8 GCDs.}
    \label{fig_scalingLUMIG}
\end{figure}

We evaluate the scaling performance of the BiCGS-GNoComm(CI) solver across multiple nodes on LUMI-G. As reported in Figure~\ref{fig_scalingLUMIG}, the solver achieves a scaling efficiency of 95\%, 95\% and 91\% for 16, 32 and 64 GCDs respectively. The efficiency decreases to 85\% at 128 GCDs and declines further with additional GPUs (65\% at 256 GCDs), as the problem size is insufficient to fully utilize the computational power of a larger number of GPUs.

To provide insights into the BiCGS-GNoComm(CI) solver performances across different hardware architectures (CPU, AMD MI250X and NVIDIA Hopper GPU), we report the time to solution for both multi-node and single-node experiments. Additionally, we break down the time spent in the computation and communication stages to offer a more detailed analysis.

\begin{figure}[!h]
    \centering
    \includegraphics[width=0.95\linewidth]{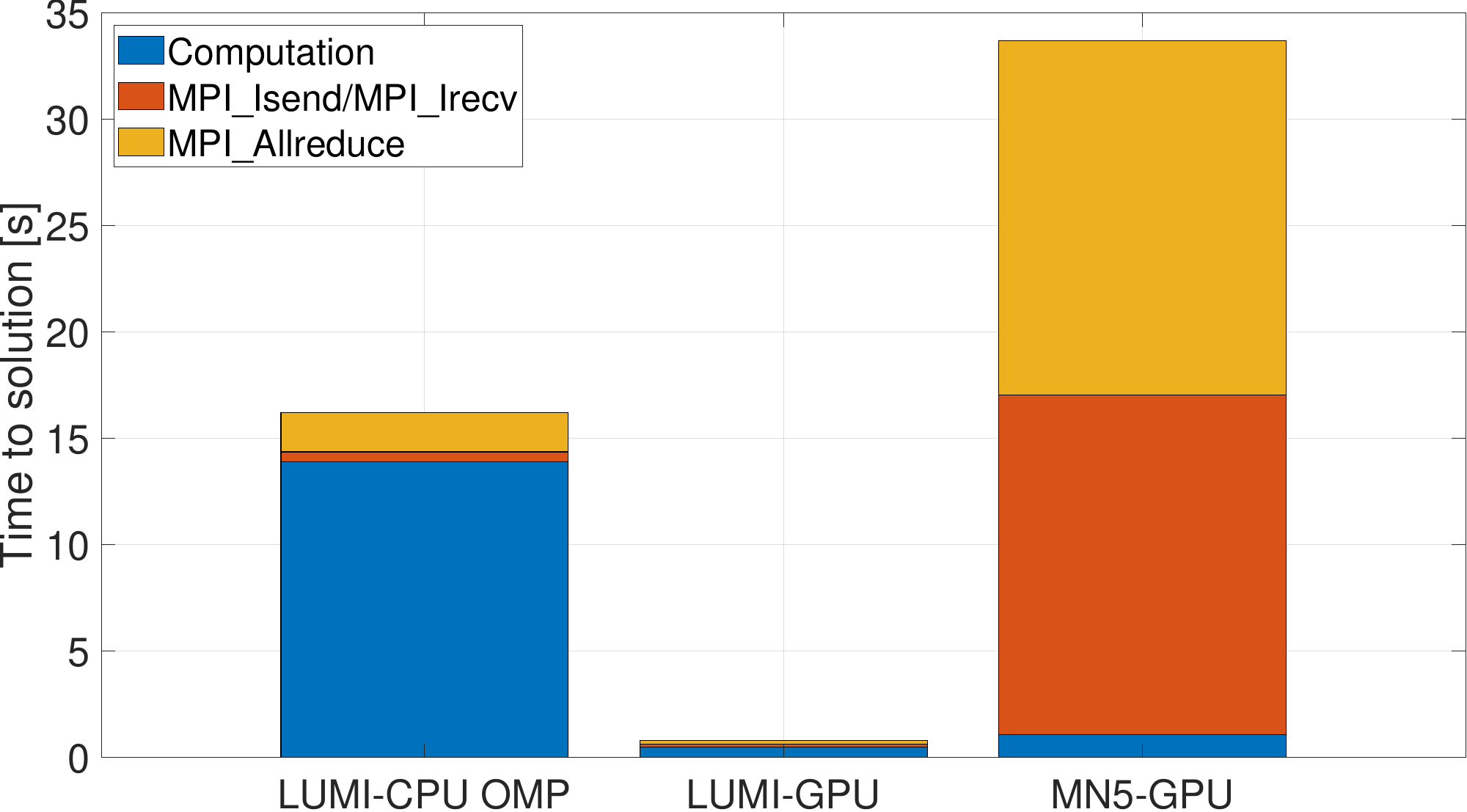}
    \caption{BiCGS-GNoComm(CI) solver time to solution across different architectures for a $256\times256\times256$ mesh and 64 MPI processes.}
    \label{fig_ttsMesh256_MPI64}
\end{figure}

\begin{figure}[h!]
    \centering
    \includegraphics[width=0.95\linewidth]{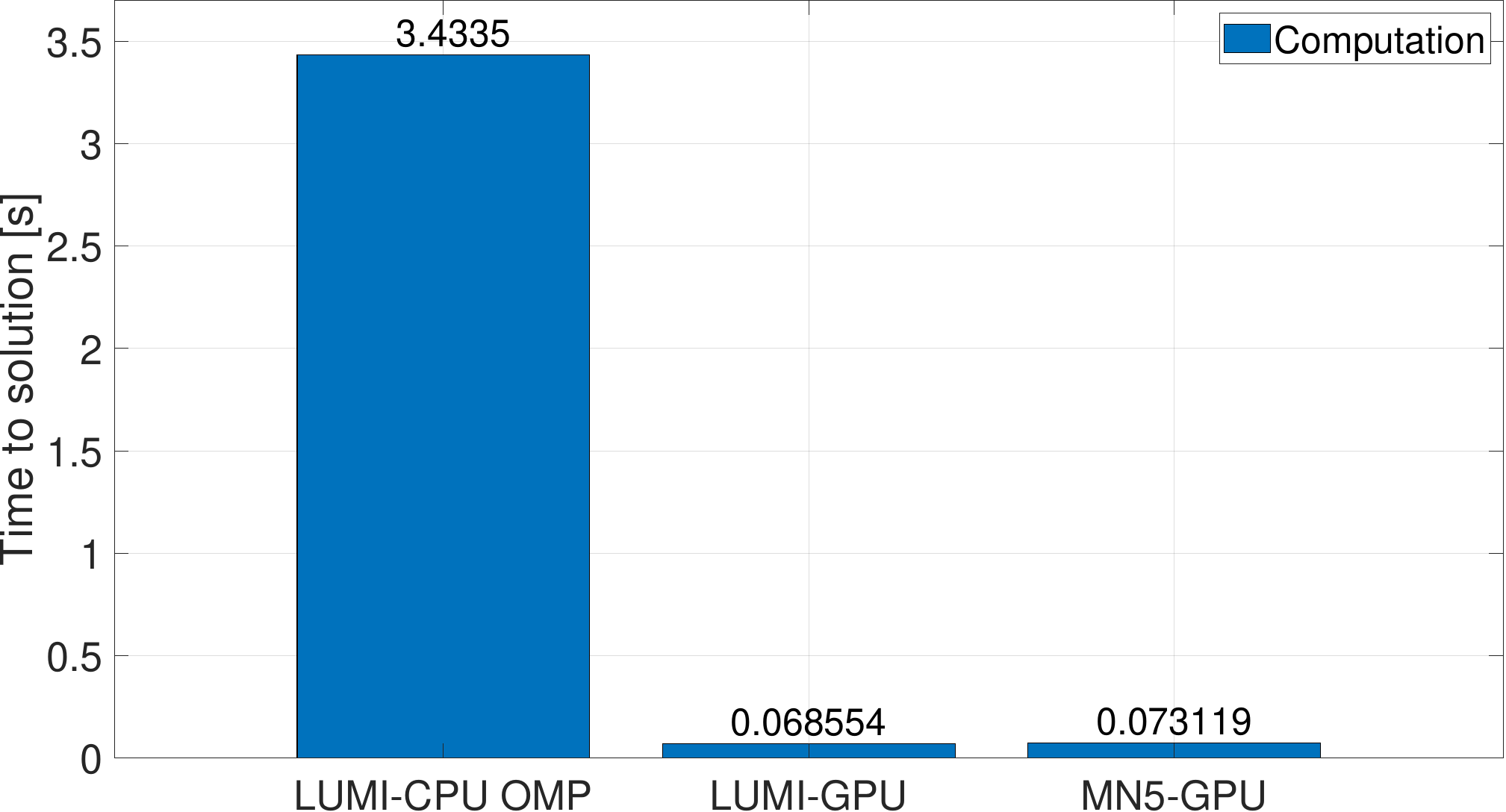}
    \caption{BiCGS-GNoComm(CI) solver time to solution across different architectures for a $64\times64\times64$ mash and one MPI process.}
    \label{fig_ttsMesh64_MPI1}
\end{figure}

Fig.~\ref{fig_ttsMesh256_MPI64} and Fig.~\ref{fig_ttsMesh64_MPI1} report the solver time to solution (averaged over 5 runs) on the different machines for the problem with $256\times 256 \times 256$ mesh run with 64 MPI ranks, and the problem with $64\times 64 \times 64$ mesh run with a single MPI rank respectively.
In both cases, the solver with AMD MI250X and NVIDIA Hopper H100 GPUs back-end significantly outperforms the CPU implementation. Specifically, in the multi-node experiment, 
AMD and NVIDIA implementations show a $29\times$ and $13\times$ faster computation compared to the CPU back-end. In the case of single node test, the GPUs achieve a computation speed-up of $50\times$ and $47\times$ for AMD and NVIDIA respectively. 

In the multi-node experiment, the NVIDIA back-end solver experiences significant delays during communication stages due to issues with the GPU-direct implementation on MareNostrum5 (Fig.~\ref{fig_ttsMesh256_MPI64}). As a result, the NVIDIA implementation is overall 42 times slower than the AMD one. For comparison, the CPU back-end solver is overall 20 times slower than the AMD implementation.

\begin{figure*}
    \centering
    \includegraphics[width=0.95\linewidth]{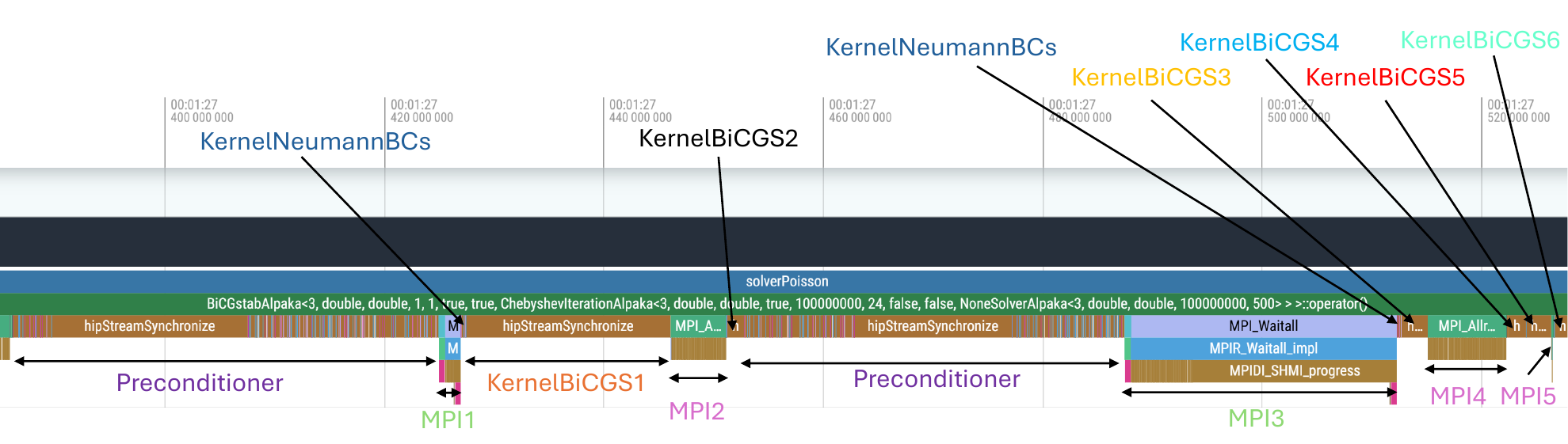}
    \caption{Traces from \texttt{Omnitrace} after running the solver on GPUs on LUMI-G: trace of one cycle of the BiCGS-GNoComm(CI) solver, Algorithm~\ref{algo_bicgstabImplementation}.} 
    \label{fig_traceLUMIGPU}
\end{figure*}

The trace output obtained during the execution of the BiCGS-GNoComm(CI) solver on LUMI-G with the AMD GPU \texttt{rocprof} and \texttt{Omnitrace} tools provides us with insights about solver performance. While we note the overhead of using a tracing impact the total solver execution time, it allows us to understand the relative impact of MPI communication and GPU kernel execution. Fig.~\ref{fig_traceLUMIGPU} showcases the annotated traces obtained from \texttt{Omnitrace}. It shows the trace for one cycle of the BiCGS-GNoComm(CI) solver (Algorithm~\ref{algo_bicgstabImplementation}), presenting the execution of the different kernels over time. When investigating the trace of one iteration of the BiCGS-GNoComm(CI) solver, we observe that the preconditioner and \texttt{KernelBiCGS1} (a matrix-vector multiplication) provide a relatively high GPU workloads, while the remaining kernels only take a small part of the AMD GPU computation. When investigating the impact of MPI communication, the largest cost is due to the MPI communication synchronization of the \texttt{MPI\_WaitAll()} call during the halo point exchange. This relatively long time is likely due of a computational imbalance present in solver execution on the distributed GPU system. We also note that the cost of the reduction operations \texttt{MPI\_AllReduce()} is not negligible.

\observation{\textit{For large problems, the solver demonstrates a strong scaling efficiency on AMD MI250X GPUs of more than $90\%$ up to 64 GCDs, and an efficiency of $85\%$ for 128 devices. The solver achieves comparable computational performances on AMD MI250X and NVIDIA Hopper H100 GPUs, both at single-node and multi node-level, greatly outperforming the corresponding CPU implementation. Specifically, AMD MI250X GPUs show a speed-up in computation compared to the CPU back-end of $50\times$ and $29\times$ for the single and multi node test respectively. The corresponding speed-up obtained with NVIDIA Hopper GPUs is $47\times$ and $13\times$, respectively.}}

\section{Related work}
\noindent \textbf{HPC Linear Solvers.} Efficient parallelization of linear solvers and effective accelerator usage is critical for solving large-scale scientific and engineering problems on supercomputers. PETSc (Portable, Extensible Toolkit for Scientific Computation) is one of the most widely used libraries for parallel scientific computing, it supports GPU execution and it is known for its scalability in supercomputer environments~\cite{mills2021toward}. Another widely used framework, Trilinos, is a collection of packages and it includes a variety of linear solvers and preconditioners~\cite{heroux2005overview}. Hypre is primarily known for its high-performance parallel multigrid preconditioners. Eigen is a high-level C++ template library for linear algebra and linear solvers~\cite{falgout2006design}.  On the GPU side, cuSPARSE provides specialized implementations of sparse matrix operations and linear solvers optimized for NVIDIA GPUs~\cite{naumov2010cusparse}. 

\noindent \textbf{HPC Frameworks for Performance Portability.} As HPC systems are evolving with diverse hardware architectures, several approaches have been developed to allow performance portability across different processing devices. Kokkos is a widely adopted C++ library that provides an abstraction layer for parallel execution and memory management on a wide variety of hardware, including CPUs and GPUs~\cite{trott2021kokkos}. RAJA is another parallel programming framework focused on providing abstractions for performance portability across diverse hardware~\cite{beckingsale2019raja}. DaCe (Data-Centric Parallel Programming) is a framework that focuses on decoupling domain science from performance optimization, enabling efficient mapping of scientific algorithms across heterogeneous hardware architectures (CPU, GPU, and FPGA) with high utilization \cite{andersson2023case, andersson2024towards}.

\section{Discussion and Conclusion}
In this paper, we presented an HPC implementation of the Bi-CGSTAB algorithm with multiple preconditioners for solving the discretized Poisson equation on HPC heterogeneous platforms. Our solver is capable of handling large-scale simulations thanks to its multi-level parallelism, in particular our approach utilizes the MPI standard for distributed memory parallelism and the \texttt{alpaka} library for shared memory parallelism. Leveraging the \texttt{alpaka} library, we can decouple the implementation from the hardware back-end, thus ensuring cross-platform portability and high optimization across several type of architectures, including CPUs, NVIDIA and AMD GPUs. This strategy allows us to have a unified codebase for all the back-ends, making the developing and maintenance of the code easier.

The results obtained in our tests on CPUs, AMD MI250X, and NVIDIA Hopper H100 GPUs, showed that all the studied preconditioning methods were effective in reducing the required BiCGSTAB solver iterations by more than eight times. The communication-free GNoComm(CI) preconditioner, based on the Chebyshev iteration, gave the lowest overall time to solution, $6.5$ faster than the unpreconditioned solver, and $50$ time faster than the FBiCGS-G(BiCGS) algorithm. The convergence properties of the BiCGS-GNoComm(CI) solver were retained across the different GPUs architectures, both at single-node and multi-node level, while, particularly in the case of the single-node test, the CPU implementation required almost twice the GPUs iterations to reach convergence (27 vs 14 iterations).

The solver demonstrated comparable computational performances on both AMD MI250X and NVIDIA Hopper H100 GPUs, while the CPU back-end appeared less optimized. Specifically, the AMD implementation showed a speed-up in computation compared to the CPU back-end of $50\times$ and $29\times$ for the single and multi node test, respectively. The corresponding speed-up obtained with NVIDIA Hopper H100 GPUs was $47\times$ and $13\times$ respectively. 
The GPU scaling test on LUMI-G across multiple nodes confirmed that the BiCGS-GNoComm(CI) solver maintains a high scaling speed-up efficiency, greater than $90\%$, up to 64 GCDs, provided the problem size is sufficiently large to fully utilize the GPU computational power.

Future work will focus on investigating the benefit of communication-avoiding/reducing algorithms to further optimize data transfer between CPU and GPU and across multi-node on supercomputers. 

\section*{Acknowledgment}
This work is funded by the European Union. This work has received funding from the European High Performance Computing Joint Undertaking (JU) and Sweden, Finland, Germany, Greece, France, Slovenia, Spain, and the Czech Republic under grant agreement No. 101093261, Plasma-PEPSC.

\ifCLASSOPTIONcaptionsoff
  \newpage
\fi

\bibliographystyle{IEEEtran}
\bibliography{IEEEabrv,IEEEexample}

\end{document}